\documentclass[italian,english,aps,manuscript]{revtex4}
\usepackage[T1]{fontenc}
\usepackage[latin9]{inputenc}
\usepackage{graphicx}
\usepackage{amssymb}
\usepackage{esint}

\makeatletter

\providecommand{\tabularnewline}{\\}

\@ifundefined{textcolor}{}
{%
 \definecolor{BLACK}{gray}{0}
 \definecolor{WHITE}{gray}{1}
 \definecolor{RED}{rgb}{1,0,0}
 \definecolor{GREEN}{rgb}{0,1,0}
 \definecolor{BLUE}{rgb}{0,0,1}
 \definecolor{CYAN}{cmyk}{1,0,0,0}
 \definecolor{MAGENTA}{cmyk}{0,1,0,0}
 \definecolor{YELLOW}{cmyk}{0,0,1,0}
 }

\makeatother

\usepackage{babel}

\begin{document}

\title{Ideal chains with fixed self-intersection rate}

\author{Simone Franchini}

\affiliation{Dipartimento di Fisica, Sapienza Università di Roma, Piazzale A.
Moro 2, I-00185 Roma, Italy}

\email{simone.franchini@yahoo.it}
\begin{abstract}
We consider ideal chains in a hypercubic lattice $\mathbb{Z}^{d}$,
$d\geq3$, with a fixed ratio $m$ of self-intersection per monomer.
Despite the simplicity of the geometrical constraint, this model shows
some interesting properties, such as a collapse transition for a critical
value $m_{c}$. Numerical simulations show a Self-Avoiding-Walk-like
behavior for $m<m_{c}$, and a compact cluster configuration for $m>m_{c}$.
The collapse seems to show the same characteristics as the canonical
thermodynamical models for the coil-globule transition.
\end{abstract}



\maketitle

\section{Introduction}

Polymer models have been the subject of extensive theoretical and
numerical studies. Most of these models are based on ideal flexible
chains, with the addition of various kinds of interactions between
monomers, to include some non-trivial properties of real systems \cite{des Cloizeaux,De Gennes-Scal.}. 

Very important is the \emph{excluded volume} effect, which significantly
modifies the fractal properties (the chains behave like Self-Avoiding
Walks, SAW); and the coil-globule (CG) transition, in which a flexible
chain collapses from an extended coil to a liquid-like globule \cite{De Gennes-Scal.,Grosberg-Kuznestov}.

Most of the thermodynamic models that show the CG transition usually
consider a competition between interactions of different geometrical
nature. For instance, one can consider a lattice random chain with
repulsive on-site interactions and attractive nearest-neighbors links.
The transition arises from the competition of these interactions.

Let $\omega=\left\{ S_{0},S_{1},\,...\,,S_{n}\right\} $ be an $n-$step
Simple Random Walk (SRW) on the $\mathbb{Z}^{d}$ lattice ($S_{i}$
are lattice vectors), and define the number of visits to each site
$x$ as $\psi_{x}=\sum_{i}\delta\left(x-S_{i}\right)$. A canonical
model, incorporating excluded volume and a CG transition, is described
by the Hamiltonian\begin{equation}
\mathcal{H}\left[\psi\right]=\epsilon_{0}\sum_{x}\psi_{x}^{2}-\epsilon_{1}\sum_{\langle x,y\rangle}\psi_{x}\psi_{y},\label{eq:0}\end{equation}
where $\langle x,y\rangle$ indicates nearest-neighbors (taking $\epsilon_{1}=0$
will lead to the Domb-Joyce model, see \cite{Huges}).

A few years ago an interesting thermodynamic model that shows a transition
with only on-site interactions was proposed \cite{Krawckzyk}. However,
a competition between two geometrically different constraints is still
present since a self-avoidance is incorporated through restricting
the maximal number of visits per site.

To our knowledge all the thermodynamic models proposed so far, having
only one kind of short-range interaction, do not allow for the CG
transition in the Boltzmann parameter.

In this work we will show how excluded volume effects, the CG transition
and liquid-like clusters can be obtained by imposing a single global
geometric constraint. We consider ideal chains in a hypercubic lattice
$\mathbb{Z}^{d}$, $d\geq3$, with a fixed ratio $m$ of self-intersections
per monomer.

This model shows a CG transition for a critical ratio $m_{c}$: numerical
simulations (obtained by a standard implementation of the Pruned-Enriched
Rosenbluth Method, PERM, see \cite{PERM,Grassberger,Prellberg,Hsu-Grassberger})
indicate a SAW-like behavior for $m<m_{c}$ (for a review about SAW
see \cite{Madras-Slade,Clisby,Owczarek}) and a cluster configuration
for $m>m_{c}$. In addition the model is amongst the simplest with
a crossover from SAW to cluster behaviour. Our focus will be on $d\geq3$
lattices because in these lattices, as we shall see, the transition
occurs at non-trivial values of $m_{c}$.

\section{Model definition}

Consider an ideal chain $\omega$ of $n$ steps on $\mathbb{Z}^{d}$
(without loss of generality we take $S_{0}=0$). We call $R\left[\omega\right]$
the range of $\omega$ (number of distinct lattice sites visited by
the path), and $M\left[\omega\right]=\left(n+1\right)-R\left[\omega\right]$
the number of self-intersections.

Let $P_{n}\left(M\right)$ be the fraction of SRW, of length $n$,
with exactly $M\in\left[0,n-1\right]$ self-intersections. We introduce
mean value $\langle M_{n}\rangle$ and variance $\langle\Delta M_{n}^{2}\rangle=\langle M_{n}^{2}\rangle-\langle M_{n}\rangle^{2}$
of the distribution $P_{n}\left(M\right)$. It is well known (see
\cite{Huges}) that for SRW in $d\geq3$, $d\neq4$,\begin{equation}
\langle M_{n}\rangle=C_{d}n-\Delta_{d}n^{2-\frac{d}{2}}+\mathcal{O}\left(1\right),\label{eq:1.1}\end{equation}
 where $C_{d}$ is the probability that an infinite length walk contains
its starting site at least twice (for numerical values of $C_{d}$,
see \cite{Douglas}), and $\Delta_{d}$ is exactly known (for $d=4$
the main fluctuation is actually $\Delta_{4}\log\left(n\right)$)
\cite{Huges,Douglas}. Concerning the variance, Jain and Pruitt have
shown that $\langle\Delta M_{n}^{2}\rangle\propto n\log\left(n\right)$
for $d=3$, and $\langle\Delta M_{n}^{2}\rangle\propto n$ for $d\geq4$
\cite{Huges,Jain-Pruitt}. They have also shown that for $d\geq2$
\begin{equation}
\xi=\lim_{n\rightarrow\infty}\left(M-\langle M_{n}\rangle\right)/\langle\Delta M_{n}^{2}\rangle\label{eq:2}\end{equation}
 is normally distributed, from which it follows that $P_{n}\left(M\right)$
is peaked around its mean value for long walks.

We are interested in chains in which the rate of intersections per
monomer is fixed at a certain value $m\in\left[0,1\right]$ in the
thermodynamic limit. We define the ensemble $\Omega_{m}$ of $n-$step
walks with exactly $M=\left\lfloor m\left(n-1\right)\right\rfloor $
intersections, and $\Omega$, the ensemble of all $n-$step SRW: we
call $\langle\cdot\rangle_{m}$ the average on $\Omega_{m}$, and
$\langle\cdot\rangle$ that of $\Omega$.

For large $n$ we can approximate $\left\lfloor m\left(n-1\right)\right\rfloor \simeq mn$.
Henceforth we will work under this approximation. We introduce the
fraction of $m-$intersection rate walks as $P_{n}\left(m\right)$.
From the properties of $P_{n}\left(M\right)$ it follows that $P_{n}\left(m\right)$
is peaked around $\langle m\rangle=C_{d}$ when $n\rightarrow\infty$.

\begin{figure}
\begin{centering}
\includegraphics[scale=0.6]{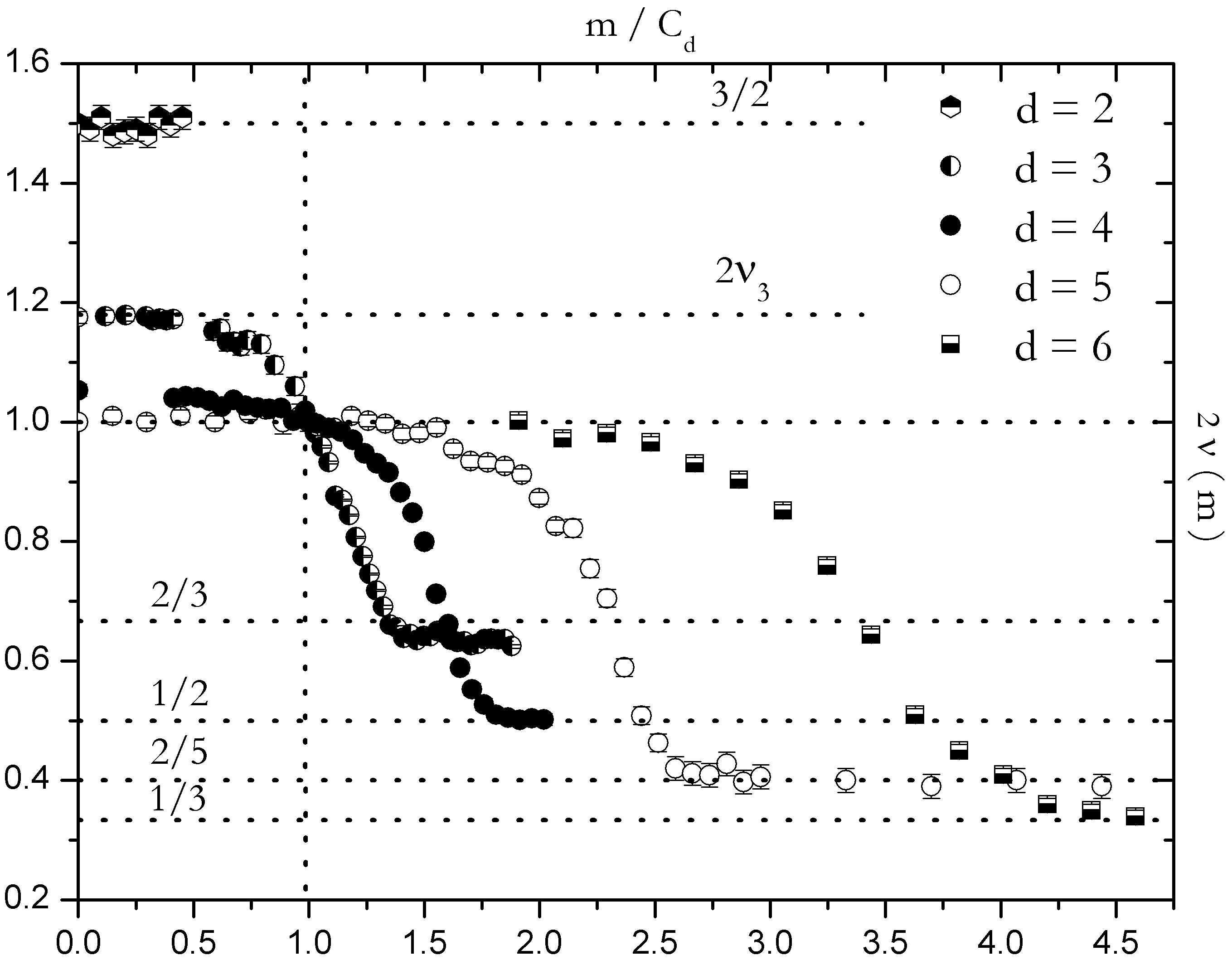}
\par\end{centering}

\caption{Exponents $\nu_{d}\left(m\right)$ vs $m/C_{d}$ from simulations
at various lattice dimensions. The graph shows the behavior of $\nu_{d}\left(m\right)$,
obtained from fits of $\log\left(\langle S_{n}^{2}\rangle_{m}\right)$
vs $\log\left(n\right)$. The range for all fits is $n\in\left[0.8\cdot n_{M},n_{M}\right]$.
Maximal length for simulated walks was: $n_{M}=2\cdot10^{3}$ for
$d=2$, $n_{M}=10^{3}$ for $d=3$, $n_{M}=10^{3}$ for $d=4$, $n_{M}=0.5\cdot10^{3}$
for $d=5$, and $n_{M}=0.3\cdot10^{3}$ for $d=6$. Moreover, for
$d=3$, the $\log\left(\langle S_{n}^{2}\rangle_{m}\right)$ has been
achieved by using $m=C_{3}M/\langle M_{n}\rangle$, with an $n$ dependence,
to take into account the finite size of the chains (see Fig. (\ref{Flo:Fig2})).
For $d>3$, the finite size effects are much weaker and this correction
is not necessary. The picture for $d=4$, $m<m_{c}$ shows an exponent
slightly larger than $1$ (a correction of order $\mathcal{O}\left(10^{-2}\right)$),
consistent with the expected logarithmic correction (for chains of
length $10^{3}$ an exponent of $2\nu\simeq1.070$ would be predicted).
In the $n\rightarrow\infty$ limit, $\nu_{d}\left(m\right)$ is expected
to be a step function (see Fig. (\ref{Flo:Fig3})). The present graph
shows the step at $m_{c}$: for $d=3,4$ we expect $m_{c}=C_{d}$,
for $d\geq5$ simulations suggest $m_{c}>C_{d}$.}

\centering{}\emph{\label{Flo:Fig1}}
\end{figure}

\begin{figure}
\centering{}\includegraphics[scale=0.6]{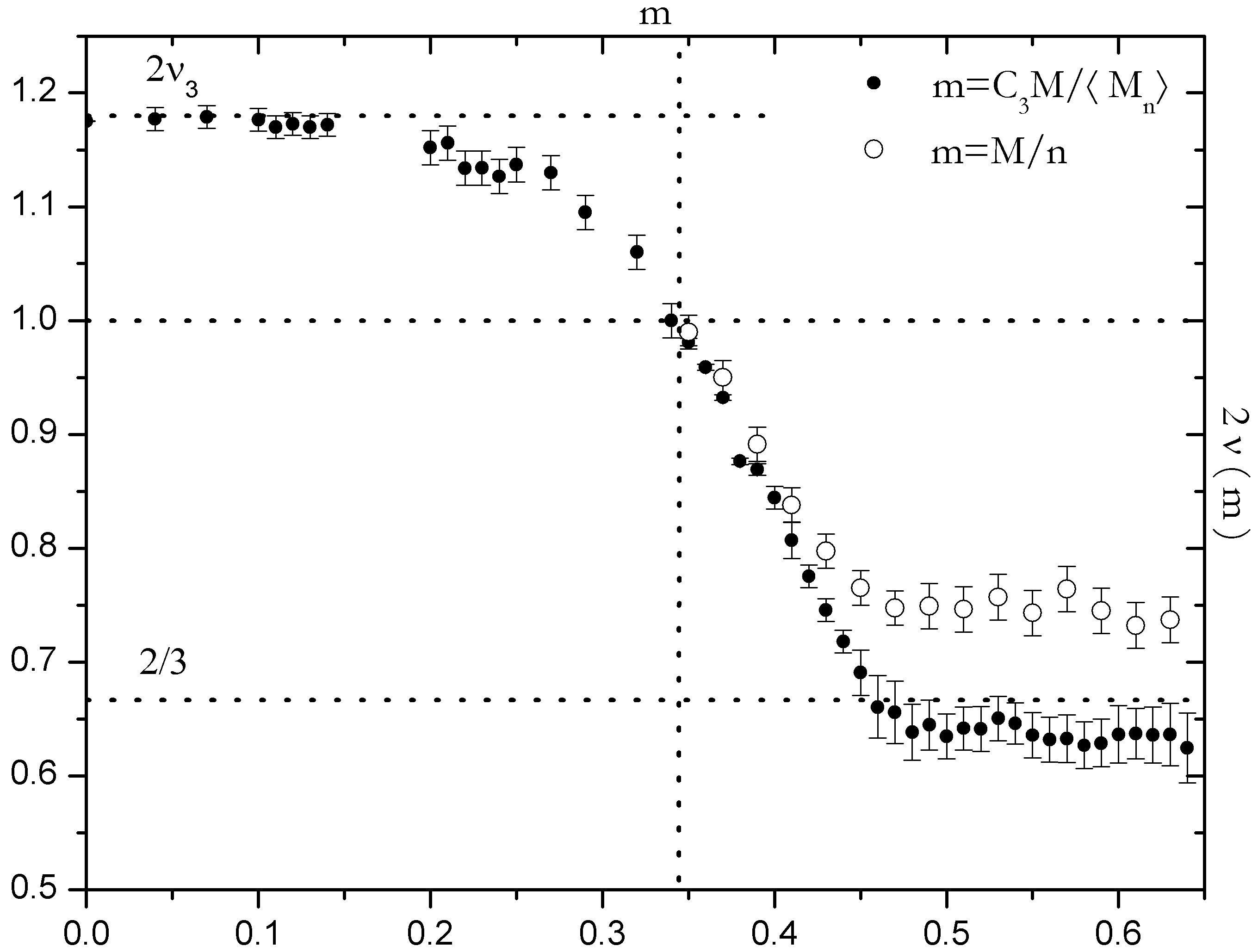}\caption{Exponent $\nu_{3}\left(m\right)$ vs $m$, a comparison of simulations
with and without finite size corrections. The chart, from fits in
the form $\log\left(\langle S_{n}^{2}\rangle_{m}\right)$ vs $\log\left(n\right)$,
compares $m=M/n$ (empty circles) with $m=C_{3}M/\langle M_{n}\rangle$
(full circles), $\langle M_{n}\rangle$ is obtained from Eq. (\ref{eq:1.1}).
Essentially, $M$ is rescaled with the average value of the support
(range) $\langle R_{n}\rangle=n-\langle M_{n}\rangle$: this procedure
leads to a significant improvement of the accuracy, at least for $m>C_{3}$
(vertical dotted line). Given that the two curves in the limit $n\rightarrow\infty$
must converge, it is clear that for $n_{M}=10^{3}$ we are still far
from the asymptotic regime. This fact is emphasized in Fig. (\ref{Flo:Fig3})
in which we can see how the \emph{drop band} for $d=3$ vanishes very
slowly compared to $d>3$. }
\label{Flo:Fig2}
\end{figure}

Let $\langle S_{n}^{2}\rangle_{m}$ be the mean square end-to-end
distance on $\Omega_{m}$: from our simulations we find that the relation
$\langle S_{n}^{2}\rangle_{m}\propto n^{2\nu_{d}\left(m\right)}$
holds for any dimension considered, with the exponent $\nu_{d}\left(m\right)$
dependent on $m$ and $d$. Again, for $d\geq3$, the simulations
show the existence of a critical value $m_{c}\in\left(0,\,1\right)$
beyond which the chains collapse into a compact liquid-like globule,
with $\nu_{d}\left(m\right)=1/d$. If instead $m<m_{c}$ we have the
SAW-like behavior (Fig. (\ref{Flo:Fig1}) and (\ref{Flo:Fig2})).
As preliminary observation we can state that, since for $m=0$ we
have the Self-Avoiding Walk, then $\langle S_{n}^{2}\rangle_{0}\simeq D_{d}n^{2\nu_{d}}$
($\nu_{d}$ is the correlation length exponent for the SAW, see \cite{Madras-Slade}).
On the other hand, from Eq. (\ref{eq:2}) it follows that the relation
\begin{equation}
\int_{-b_{n}}^{b_{n}}d\epsilon\,\langle S_{n}^{2}\rangle_{C_{d}+\epsilon}P_{n}\left(C_{d}+\epsilon\right)\simeq n,\label{eq:3.1}\end{equation}
should hold in the large $n$ limit and $b_{n}=o\left(1\right)$.
This is clearly confirmed by our simulations.

\begin{figure}
\centering{}\includegraphics[scale=0.6]{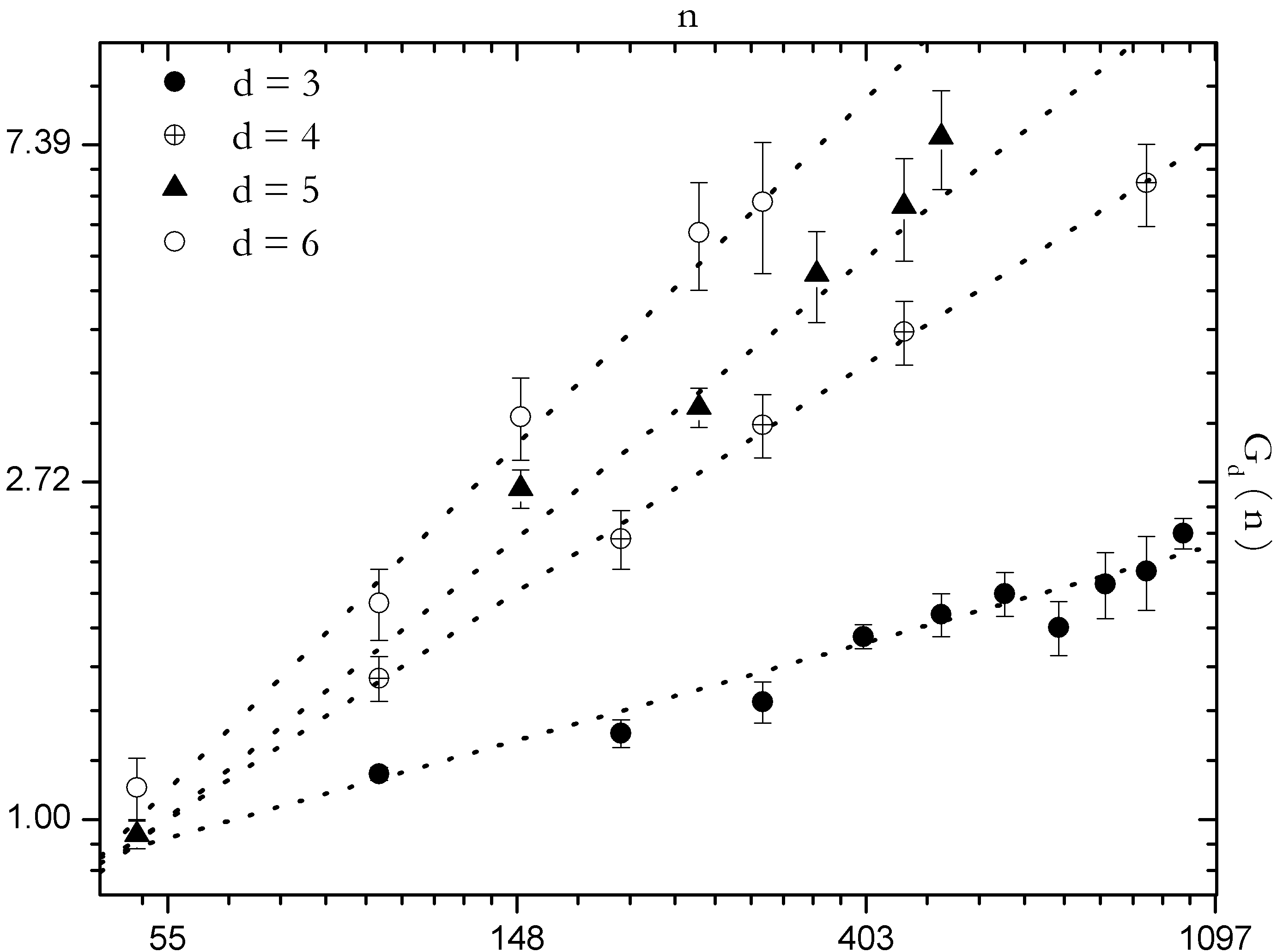}\caption{Slope of $\nu_{d}\left(m\right)$ in the drop band. The graph shows
(in log-log scale) the maximum slope $G_{d}\left(n\right)=-\left[\partial_{m}\nu_{d}\left(m\right)\right]_{min}$
of $\nu_{d}\left(m\right)$ for different lengths $n$, which is an
estimator of the \emph{drop band} width (proportional to $G_{d}^{-1}\left(n\right)$)
in which $\nu_{d}(m)$ falls from $\nu_{d}$ to $1/d$. $G_{d}\left(n\right)$
seems to increase with a power law, supporting the idea that $\nu_{d}\left(m\right)$
is a step function in the limit $n\rightarrow\infty$. Dotted lines
are power law $B\cdot n^{\alpha}$ fits to the data: $\alpha=0.29\pm0.01$
for $d=3$ (without the finite size correction $C_{3}/\langle M_{n}\rangle$),
$\alpha=0.67\pm0.03$ for $d=4$, $\alpha=0.83\pm0.05$ for $d=5$,
and $\alpha=1.02\pm0.07$ for $d=6$. From this data we can notice
that the width of the \emph{drop band} decreases faster in $n$ as
lattice dimension $d$ increases.}
\label{Flo:Fig3}
\end{figure}

\section{General results}

We will briefly discuss the recurrent cases $d=1,2$. For $d=1$ we
have a rather simple situation: in a linear lattice chain the end-to-end
distance is proportional to the range, hence proportional to $n$,
$\forall m<1$. A unidimensional SRW is recurrent ($C_{1}=1$, see
\cite{Huges}), therefore no drop of $\nu_{1}\left(m\right)$ is expected
for $m<1$. At $d=2$ the picture is conceptually similar. We find
$\nu_{2}\left(m\right)=\nu_{2}$ (data in Fig. (\ref{Flo:Fig1})),
and since the square lattice SRW is still recurrent ($C_{2}=1$, from
\cite{Huges}) from Eq. (\ref{eq:3.1}) we expect that this relation
holds for any $m<1$ (see later discussion about the case $d=3$).

The most important case is clearly $d=3$: from simulations we find
that $\nu_{3}\left(m<m_{c}\right)=\nu_{3}$, and $\nu_{3}\left(m>m_{c}\right)=1/3$.
The last statement needs some attention, we point out that $\nu_{3}\left(m>m_{c}\right)=1/3$
is reached only after a long crossover. Fig. (\ref{Flo:Fig2}) shows
$\nu$ vs $n^{-1}M$ and $C_{3}M/\langle M_{n}\rangle$: the former
does not consider the finite size of the system, while the latter
takes into account fluctuations of the walk mean support (range) $\langle R_{n}\rangle$
at finite $n$ (see \cite{Franchini-0}). This enhancement allows
a significant improvement of the accuracy for $d=3$ (while is useless
in higher dimensions, since, by Eq. (\ref{eq:1.1}), the range converges
rapidly to its asymptotic behavior).

From Eq. (\ref{eq:3.1}) we can locate $m_{c}$ for $d=3$. If $\langle S_{n}^{2}\rangle_{m}$
has the SAW-like behavior (with $\nu_{3}>1/2$, see \cite{Clisby})
for $m<m_{c}$, and cluster-like for $m>m_{c}$, then the drop of
$\nu_{3}\left(m\right)$ must lie within the range of integration
of Eq. (\ref{eq:3.1}). It follows that $m_{c}=C_{3}$ for $n\rightarrow\infty$.
As for the exponent at the critical point, Eq. (\ref{eq:3.1}) tells
us that $\nu_{d}\left(m\rightarrow C_{d}\right)=1/2$, from which
it follows that $\nu_{3}\left(m_{c}\right)=1/2$: this observation
is supported by Fig. (\ref{Flo:Fig1}) and (\ref{Flo:Fig2}), where
we see that $2\nu_{3}\left(m\right)$ passes through $1$ at the expected
critical point $m_{c}=C_{3}$.

This latter fact is of some importance, since the transition for $d=3$
would show the same behavior as that described by the Hamiltonian
in Eq. (\ref{eq:0}). If the two models belong to the same universality
class this would be very interesting, since in our model the transition
arises from the necessity of maximizing the configurational entropy,
without the need for any further interactions. Indeed, at the critical
point $m_{c}$ there is a radical change in the optimal strategy to
achieve the global constraint $m$ (which is actually a long-range
correlation): for $m<m_{c}$, the best way to change the ratio $m$
is to compress (or expand) the chain locally, keeping the SAW-like
fractal structure. In this situation the monomers intersect (on average)
only within a certain distance along the chain. For $m>m_{c}$ instead,
it becomes entropically convenient to assume a compact configuration,
also allowing intersections between monomers very far apart in terms
of position along the chain.

We also studied higher dimensions, the results are consistent with
a CG transition of the same kind. For $d=4$ we have $\nu_{4}\left(m>m_{c}\right)=1/4$
and $\nu_{4}\left(m<m_{c}\right)=1/2$, but the random-coil behavior
shows logarithmic corrections. Simulations fit the conjecture that
$\langle S_{n}^{2}\rangle_{m}\propto n\log\left(n\right)^{1/4}$,
as in the SAW case \cite{Madras-Slade}. If this is true, from Eq.
(\ref{eq:3.1}), we should again find $m_{c}=C_{4}$.

As expected, for $d\geq5$, $m<m_{c}$, we find the mean field behavior
$\nu_{d}\left(m\right)=1/2$ (as for the SRW), and $\nu_{d}\left(m\right)=1/d$
for $m>m_{c}$. Since for $d\geq5$ $\nu_{d}\left(m<m_{c}\right)$
is the same as for the SRW, we neither use Eq. (\ref{eq:3.1}) to
locate $m_{c}$ nor find $\nu_{d}\left(m_{c}\right)$, but our simulations
(extremes of the drop zone of $\nu_{d}\left(m\right)$, $d=5,6$ in
Fig. (\ref{Flo:Fig1})) strongly suggest that $m_{c}>C_{d}$. All
results and conjectures about the behavior of $\langle S_{n}^{2}\rangle_{m}$
have been summarised in Table I.

\begin{table}
\selectlanguage{italian}%
\centering{}\begin{tabular}{|c|c|c|c||c|}
\hline 
\noalign{\vskip\doublerulesep}
$d$ & $\nu\left(m<m_{c}\right)$ & $\nu\left(m=m_{c}\right)$ & $\nu\left(m>m_{c}\right)$ & $m_{c}/C_{d}$\tabularnewline[\doublerulesep]
\hline
\hline 
$3$ & $\nu_{3}$ & $1/2$ & $1/3$ & $1$\tabularnewline
\hline 
$4$ & $\,\,\,\,\,\,\,\,1/2\,_{l.c.}$ & $1/2$ & $1/4$ & $1$\tabularnewline
\hline 
$5$ & $1/2$ & $-$ & $1/5$ & $1.5\backsim2.7$\tabularnewline
\hline 
$6$ & $1/2$ & $-$ & $1/6$ & $2.1\backsim4.4$\tabularnewline
\hline
\end{tabular}\caption{\selectlanguage{english}%
Summary of predictions for $\nu_{d}\left(m\right)$: for $d=5,6$\emph{
}we where unable to look at the critical behavior, while the critical
point $m_{c}$ is evaluated by the drop zone of $\nu_{d}\left(m\right)$.
The\emph{ l.c.} indicates SAW logarithmic correction in $d=4$.\selectlanguage{italian}
}
\selectlanguage{english}

\end{table}

We would like to point out that all results for $d\geq3$, $m\geq m_{c}$,
are consistent with a remarkable exact work by M. van den Berg, E.
Bolthausen and F. den Hollander (\cite{den Hollander}) on the moderate
deviations for the volume of a Wiener Sausage (WS), which is a neighborhood
of the trace of a standard Brownian motion up to a time $t$, given
by taking all points within a fixed distance $a$ of Brownian motion
(essentially a continuous version of our model, see \cite{Franchini-1}).
Let $\eta\left(t\right)$, $t\geq0$ be the standard Brownian motion
in $\mathbb{R}^{d}$ starting at the origin. The WS $W^{a}\left(t\right)$
with radius $a$ is the process defined by \begin{equation}
W^{a}\left(t\right)=\bigcup_{0\leq s\leq t}B_{a}(\eta\left(s\right)),\end{equation}
where $B_{a}\left(x\right)$ is the open ball with radius $a$ around
$x\in\mathbb{R}^{d}$. This paper considers the probability of having
a Wiener Sausage $W^{a}\left(t\right)$ of volume $\left|W^{a}\left(t\right)\right|\leq bt$,
$b\in\left[0,\kappa_{a}\right]$ ($\kappa_{a}t$ is the long time
behavior of $\langle\left|W^{a}\left(t\right)\right|\rangle$), showing
that there exists a critical value $b_{c}$ below which the sausage
is supposed to collapse in a \emph{swiss-cheese} like compact configuration
(a non-percolating cluster with random holes of size $\mathcal{O}\left(1\right)$).
They rigorously showed that $b_{c}\in\left(0,\kappa_{d}\right)$ for
$d\geq5$ only, while for $d=3,4$ the transition is at $b_{c}=\kappa_{a}$
exactly.

\section{Order parameter}

Typically the order parameter considered for a CG transition is $n^{-1}\langle S_{n}^{2}\rangle_{\beta}$
(with $\langle\cdot\rangle_{\beta}$ being the thermodynamic average);
we propose here a slightly different parameter: \begin{equation}
\varrho_{d}\left(m\right)=\lim_{n\rightarrow\infty}\sqrt{\langle S_{n}^{2}\rangle_{m}/\langle S_{n}^{2}\rangle_{0}}.\label{eq:4}\end{equation}
 This function vanishes beyond the critical point, similarly to the
magnetization for spin systems. If the analogy with the CG transition
from Eq. (\ref{eq:0}) holds, we should expect a second order transition,
at least for $d=3$.

\begin{figure}
\centering{}\includegraphics[scale=0.6]{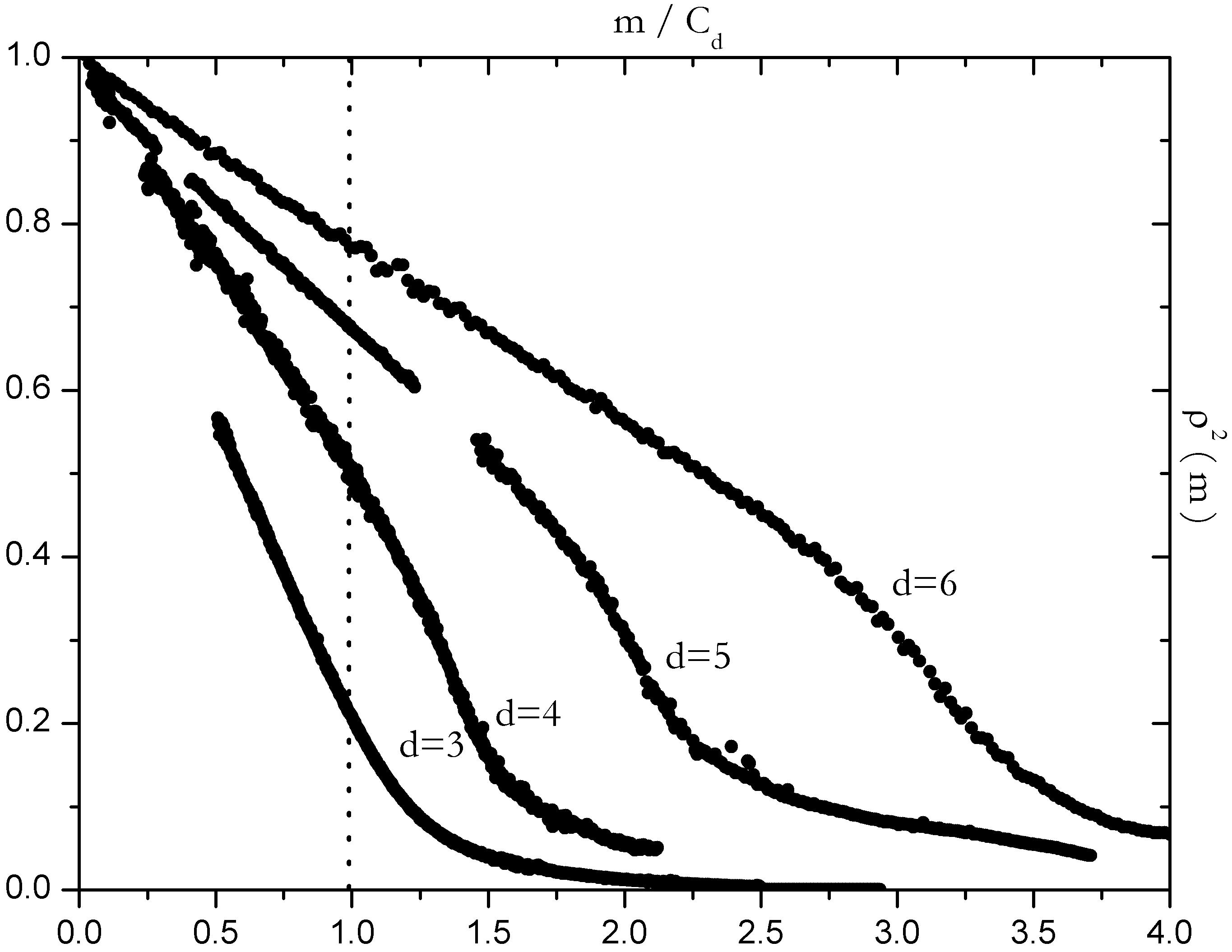}\caption{Order parameter $\varrho_{d}\left(m\right)$ vs $m/C_{d}$ from simulations
at various lattice dimensions. The graph shows $\langle S_{n}^{2}\rangle_{m}/\langle S_{n}^{2}\rangle_{0}$:
$n=10^{3}$ for $d=3$, $n=2\cdot10^{3}$ for $d=4$, $n=10^{3}$
for $d=5$ and $n=0.5\cdot10^{3}$ for $d=6$. The vertical dotted
line is the expected transition point for $d=3,4$ as $n\rightarrow\infty$.
The asymptotic behavior near $m_{c}$ is only hinted, since our PERM
implementation, which requires us to store the microcanonical density
of states at many $n$, has not allowed us to simulate longer chains.
Better pictures could surely be achieved, by more focused techniques,
for every dimensions except $d=4$. For $d=4$ and $m<C_{4}$ we expect
$n^{-1}\langle S_{n}^{2}\rangle_{m}$ to be only $\mathcal{O}(\log\left(n\right)^{1/4})$,
implying the asymptotic behavior would be difficult to demonstrate
clearly even by simulation for very large chains.}
\label{Flo:Fig4}
\end{figure}

Our simulations (Fig. (\ref{Flo:Fig4})) do not allow us to clarify
if the transition is continuous at $m_{c}$. However, for $d=3$ we
can get some insights by a mean field analysis of the Stanley Model
(SM, see \cite{Franchini-2}), a model of correlated random walks
defined by the following partition function: \begin{equation}
\mathcal{Z}_{n}\left(\beta\right)\propto\sum_{M=0}^{n}P_{n}\left(M\right)e^{-\beta M}.\label{eq:5}\end{equation}
 Given that variable $\xi$ in Eq. (\ref{eq:2}) is normally-distributed,
it's easy to show that for small $\beta$ \begin{equation}
\lim_{n\rightarrow\infty}n^{-1}\langle M_{n}\rangle_{\beta}\simeq C_{3}-\delta m\left(\beta\right),\label{eq:6}\end{equation}
 where $\langle M_{n}\rangle_{\beta}$ is the mean number of intersections
in the SM at temperature $\beta$, and $\delta m\left(\beta\right)$
vanishes for $\beta\rightarrow0$. From Eq. (\ref{eq:2}) we have
a Gaussian shape $P_{n}\left(M\right)$ near $M=C_{3}n$, with $\langle\Delta M_{n}^{2}\rangle\propto n\log\left(n\right)$.
By replacing $M\rightarrow mn$ in Eq. (\ref{eq:5}) we find \begin{equation}
\mathcal{Z}_{n}\left(\beta\right)\propto\int_{0}^{1}dm\, e^{-\frac{n}{2}\left[2\beta m+n\langle\Delta M_{n}^{2}\rangle^{-1}\delta m^{2}\right]}\label{eq:7}\end{equation}
for small $\beta$, with $\delta m=\left(C_{3}-m\right)\simeq0$.
Whereas there is a positive constant $a_{0}$ such that $\langle\Delta M_{n}^{2}\rangle>a_{0}n$
for large $n$, performing a saddle point integration on Eq. (\ref{eq:7})
with $n\rightarrow\infty$ we obtain $\delta m\left(\beta\right)>c_{0}\beta$
(for small $\beta$ and $c_{0}>0$).

From simple mean field arguments (see \cite{Franchini-3}) it is reasonable
to assume that the SM can be approximately described by the Flory
theory. Consider the following Flory energy\begin{equation}
\mathcal{F}_{\beta}\left(r\right)\sim\beta n^{2}r^{-d}+\frac{r^{2}}{2n}-\left(d-1\right)\log\left(r\right)+\Lambda\left(n\right),\label{eq:8}\end{equation}
where $r=\left|S_{n}\right|$ is the end-to-end distance, and $\Lambda\left(n\right)$
is a function independent of $r$ \cite{Madras-Slade}. We minimize
the functional under the assumption $r\thicksim\beta^{\theta}n^{\nu}$,
thus \begin{equation}
d\beta^{1-\theta\left(1+d\right)}n^{2-\nu\left(1+d\right)}+\left(d-1\right)\beta^{-\theta}n^{-\nu}=\beta^{\theta}n^{\nu-1}.\label{eq:9}\end{equation}
When $\nu d<2$ we find $\nu=3/\left(d+2\right)$ and $\theta=1/\left(d+2\right)$,
otherwise $\nu=1/2$ and $\theta=0$. When $d=3$ this theory predicts
$\theta=1/5$. A different estimation method uses path integrals,
results are in agreement \cite{Kleinhert}. 

Flory theory for SM where $d=3$ predicts $\langle S_{n}^{2}\rangle_{\beta}\propto\beta^{2/5}n^{6/5}$
for small $\beta$: inverting $\delta m\left(\beta\right)$, and substituting
in the expression for $\langle S_{n}^{2}\rangle_{\beta}$, we find
$\varrho_{3}\left(m\right)<b_{0}\delta m^{1/5}$, with $\delta m=\left(C_{3}-m\right)\simeq0$,
$b_{0}>0$.

In the case $d=4$ Flory theory cannot be used, since the logarithmic
correction to the mean field behavior are not captured. Note that
the $d\geq5$ theory predicts $\theta=0$, in agreement with the prediction
$m_{c}>C_{d}$ for $d\geq5$.

The exponents found are certainly incorrect, since $\nu_{3}\neq3/5$,
but a power law behavior $\varrho_{3}\left(m\right)\propto\delta m^{\theta}$
(with a likely logarithmic correction) is reasonable, suggesting that
in $d=3$ the transition is of the second order, with $m_{c}=C_{3}$.

\section{Conclusion and outlook}

In this paper we introduced a new athermal model of interacting random
walks, which shows a CG transition for a critical ratio between the
range and the number of monomers. The transition seems to show the
same characteristics as that seen in canonical models. However, the
relationship between our microcanonical model and those canonical
for the CG transition is non-trivial (for a general discussion on
the differences between microcanonical and canonical ensembles see
\cite{Touchette}).

The Hamiltonian in Eq. (\ref{eq:0}), as well as the interaction described
in reference \cite{Krawckzyk}, operate a selection on the ensemble
$\Omega$ whose mechanisms are not easily connected to the ensemble
$\Omega_{m}$. Although this is a key issue, we do not discuss it
here: a work dedicated to this subject is currently under preparation,
where we will present a detailed study of the $P_{n}\left(m\right)$
distribution and its relation to some thermodynamic models.

Apart from more focused implementation of the simulations here presented,
other issues of interest could affect the connectivity properties
of collapsed clusters (an example is found in \cite{den Hollander},
collapsed WS clusters should be non-percolating, with $\mathcal{O}\left(n\right)$
holes sized $\mathcal{O}\left(1\right)$ distributed inside the range:
is reasonable to expect the same behavior for our model when $m>m_{c}$).

In general, this model certainly deserves attention since, in our
opinion, it could lead to substantial improvements in understanding
the geometry of CG transitions, as well as the crossover between the
SAW and the SRW.

\section{Acknowledgments}

We thank Frank den Hollander (University of Leiden) for suggesting
the finite-size correction to improve simulations in $d=3$, and Jack
F. Douglas (NIST) for fundamental clarifications about the current
theory of the CG transition. We also thank Riccardo Balzan (EPFL)
and Giorgio Parisi (Sapienza Università di Roma) for interesting discussions.


\begin{thebibliography}{23}
\bibitem{Franchini-0}The optimal correction would actually be $m=1-\bar{C}_{3}R/\langle R_{n}\rangle$
instead of $m=C_{3}M/\langle M_{n}\rangle$ but, since they differ
only by $\mathcal{O}(n^{-1})$ compared to a leading order of $\mathcal{O}(n^{-1/2})$,
we preferred to use the second option to simplify the data analysis.
\bibitem{Franchini-1}The volume of a Wiener Sausage is the continuous
analogue of the range in SRW: the condition $\left|W^{a}\left(t\right)\right|\leq bt$,
$b\in\left[0,\kappa_{a}\right]$ is equivalent to $R\left[\omega\right]\leq\left(1-m\right)n$,
$m\in\left[0,C_{d}\right]$ (then $M\left[\omega\right]\geq mn$)
in our model. For a brief review on the Wiener Sausage see \cite{den Hollander}.
\bibitem{Franchini-2}The Stanley Model (SM, see \cite{Stanley} for
the original formulation, or \cite{Huges} for a review) is a model
of correlated random walks which describes polymer chains with either
repulsion or attractions. The attractive case is mathematically equivalent
to the Rosenstock Trapping Model (random walk on lattice with randomly
distributed irreversible traps: a walker encountering a trap is killed
there, see \cite{Huges}) and has been exactly solved by M. D. Donsker
and S. R. S. Varadhan \cite{Donsker-Varadhan}.
\bibitem{Franchini-3}We show that for small $\beta$ the SM is described
by the same mean field theory of the Domb-Joyce Model (see \cite{Huges}
for a review of this important model), therefore by the Flory Theory.
We start from the Domb-Joyce Hamiltonian $\mathcal{H}_{DJ}=\epsilon\sum_{x}\psi_{x}^{2}=\epsilon\sum_{x}R_{x}\psi_{x}^{2}$,
with $R_{x}=\lim_{\gamma\rightarrow0}\psi_{x}^{\gamma}$ and $\psi_{x}$
the same of Eq. (\ref{eq:0}): approximating $\psi_{x}\rightarrow\bar{\psi}=nR^{-1}$
we have $\mathcal{H}_{DJ}\sim\epsilon R\bar{\psi}^{2}$, where $\bar{\psi}$
is the average occupation of a visited site, and $R=\sum_{x}R_{x}$
is the range. It follows that $\mathcal{H}_{DJ}\sim\epsilon n\left(1-m\right)^{-1}$,
while the Hamiltonian of the SM is $\mathcal{H}_{SM}=-\epsilon nm$.
For small $\beta$ we have to consider $m=C_{d}-\delta m$, and substituting
this expression in $\mathcal{H}_{DJ}$, $\mathcal{H}_{SM}$ we find
$\delta\mathcal{H}_{DJ}\propto\delta\mathcal{H}_{SM}$, from which
it follows that the two models are equivalent in mean field description,
with different couplings $\epsilon$.
\end{thebibliography}
\end{document}